\documentclass[a4paper,amsmath,amssymb]{jpconf}
\usepackage{graphicx}
\usepackage{amsmath, amsthm, amssymb}
\begin{document}
\title{Simulating Quantum Dynamics with\\ Entanglement Mean Field Theory}

\author{Aditi Sen(De) and Ujjwal Sen}

\address{Harish-Chandra Research Institute, Chhatnag Road, Jhunsi, Allahabad 211 019, India}

\ead{aditi@hri.res.in, ujjwal@hri.res.in}

\begin{abstract}

Exactly solvable many-body systems are few and far between, and the utility of approximate methods cannot be overestimated. Entanglement
mean field theory is an approximate method to handle such systems. While mean field theories reduce the many-body system to an effective 
single-body one, entanglement mean field theory reduces it to a two-body system. And in contrast to mean field theories where the self-consistency 
equations are in terms of single-site physical parameters, those in entanglement mean field theory are in terms of both single- and two-site 
parameters. Hitherto, the theory has been applied to predict properties of the static states, like ground and thermal states, of many-body systems. 
Here we give a method to employ it to predict properties of time-evolved states. The predictions are then compared 
with known results of 
paradigmatic spin Hamiltonians.

\end{abstract}

\section{Introduction}
\label{sec:Introduction}

Strongly interacting systems, apart from dealing with  the properties of magnetic materials, have been enriched by 
several important discoveries during the last decades including high temperature  superconductivity, quantum Hall effects, etc. 
A major breakthrough in many-body physics came with the experimental realizations of such systems artificially, for example by using ultracold atoms in optical lattices  and ion traps \cite{amaderadp}. 
On the other hand, quantum information science have demonstrated an enormous utility in applications in different disciplines, ranging from communication processes to 
computer science \cite{Nielsen-Chuang}.  In the last few years, a strong connection between quantum information and many-body physics 
has been developed \cite{amaderadp, fazioRMP}. 
This includes the study of fundamental properties of many-body systems from the perspective 
of quantum information, and also the research aimed at an implementation of a quantum computer and other 
quantum information processing devices in realizable many-body systems, e.g. in ultracold atoms and ions. 

One of the main obstacles in such investigations 
is
the lack of analytical methods for handling many-body systems. 
Indeed 
only a few interacting many-body systems  can be exactly diagonalized \cite{exact, exact1}.  
Approximate and numerical methods therefore play a crucial role in studying 
many-body physics \cite{approximate}. There are several approximate methods that have been successfully used to evaluate, characterize, and understand
 the properties of model many-body systems.
Among them,  the mean field  theory (MFT), 
first introduced by P. Weiss in 1907, and several 
improvements of that theory, have become an important method for understanding  many-body physics, including phase diagrams, critical exponents, etc.  
Typically, a mean field theory reduces the parent many-body Hamiltonian into a single-body one, with the latter containing single-site physical parameters 
as undetermined variables. These undetermined single-site parameters are subsequently obtained by solving 
certain self-consistency equations, and a distinguishing property of mean field-like theories is a (or a set of) self-consistency equations 
written in terms of single-site physical parameters, like magnetization \cite{mfprime, mf, mfprimeprimeprime, mfprimeprime, mfprime1, approximate}. 
Such mean field theories, although useful in several applications, 
are not able to properly capture the two-body properties (of the parent many-body Hamiltonian) -- essential ingredients in quantum information science \cite{Nielsen-Chuang}.  

Entanglement mean field theory (EMFT) is an approximate method for studying two-site as well as single-site physical quantities, and critical phenomena of 
many-body systems \cite{emft1}. A characteristic trait of EMFT is that it deals with self-consistency equations in terms of two-body physical properties, in contrast to
the self-consistency equations in terms of single-site quantities in mean-field theories. The effort required to solve the consistency equation in EMFT is almost
or exactly the same as in MFT. 
Previous works have demonstrated that EMFT can faithfully signal the onset of quantum 
fluctuations-driven phase transitions as well as  temperature-driven phase transitions in a large variety of paradigmatic quantum 
many-body models. Moreover, the two-site as well as single 
site physical parameters as predicted by EMFT are qualitatively as well as quantitatively better than the MFT predicted ones. In Ref. \cite{classicalamader}, an EMFT-inspired
 theory was applied to classical many-body systems, and it was shown that the critical points predicted by EMFT are significantly better than the MFT-predicted values. 
EMFT was generalized to a cluster EMFT in Ref. \cite{classicalamader}, 
and was shown to provide a significant improvement over
the predictions from cluster MFT. 

However, both the previous works utilizes EMFT to understand the properties of \emph{static} systems, viz., the ground state or the equilibrium states of time-independent 
Hamiltonians. Here we apply EMFT to study the the dynamics of a many-body system, as it evolves in time, where the evolution is described by a time-dependent Hamiltonian. 
 In particular, the EMF method is applied to explore the properties of entanglement in the dynamics of the nearest neighbor quantum transverse 
Ising model. The recipe for the quantum Ising model
can also be applied to other interacting spin systems. 
Exact results are few and far between for many-body systems, even in one-dimension (1D). However, in case of the nearest neighbor one-dimensional 
quantum transverse Ising model, an exact analytical treatment is possible for the dynamics (and statics), and several important physical parameters, including 
two-site ones can be exactly obtained in closed form (see Refs. \cite{Barrouch-McCoy, Barrouch-McCoy1} and references therein).
It is therefore possible to compare the EMFT predictions in this case with the exact results, and we do so for the nearest neighbor entanglement in this system.
Previous work had shown that the EMFT prediction for the nearest neighbor entanglement matches well with the general 
behavior obtained from exact calculations, in the case of 
the ground state and  thermal states of this model \cite{emft1}. Here, we make the comparison 
for the time-evolved state.

\section{The model}

Consider the ferromagnetic quantum transverse Ising model (TIM) with nearest neighbor 
interactions on a lattice of arbitrary dimension and arbitrary geometry. The model is described
 by the Hamiltonian
\begin{eqnarray}
H_{TIM} = - J \sum_{\langle {\vec{i}}{\vec{j}} \rangle}\sigma_x^{\vec{i}} \sigma_x^{\vec{j}} - h(t) \sum_{\vec{i}} \sigma_z^{\vec{i}}, 
\label{IsingH}
\end{eqnarray}
where \(\sigma_i, i=x,y,z\) are the Pauli spin-1/2 matrices, \(J>0\) is the coupling strength, and \(h(t)\) is a time-dependent transverse magnetic field. 
We will be interested in the time-evolution of the system, for which we assume that the initial state is the canonical equilibrium state of the system at 
time \(t=0\). 
We will only consider cases where the Hamiltonian has nearest neighbor interactions, as indicated by the notation \(\langle {\vec{i}}{\vec{j}} \rangle\)
in the first summation in Eq. (\ref{IsingH}), although the considerations here carry over to more general cases. 
Since 
the field considered is time-dependent, the ensuing dynamics will be non-trivial. 
 In this paper, the time-dependence  of the magnetic field is assumed to be as follows:
\begin{eqnarray}
\label{field}
 h(t) &=& a, \quad t\leq 0, \nonumber \\
    &=& 0, \quad t>0,
\end{eqnarray}
where \(a\neq 0\). The time-evolution of the system will therefore be a response to the initial ``kick'' given to the sytem at zero time.

The one-dimensional version of \(H_{TIM}\)  can be exactly diagonalized by the Jordan-Wigner transformation \cite{Barrouch-McCoy, Barrouch-McCoy1}.
The exact spectrum can be determined. Furthermore, it is possible to obtain several physically relevent quantities of the evolved state, 
where the time evolution starts off from the canonical equilibrium state, for a wide range of time-dependent transverse fields \(h\), including  
the field that has been considered here. 

However, such investigations are not possible in higher dimensional systems. The ground states of even the time-independent versions of 
the same quantum model in 
higher dimensions, including square and triangular lattices in two-dimensions, are not known exactly.
Therefore, approximate methods or numerical simulations play an important role to study such higher dimensional systems. Numerical simulations \cite{book} include 
exact finite-size diagonalizations, density matrix renormalizations \cite{Schollwock}, quantum monte carlo \cite{qmontecarlo}, etc., 
while  approximate methods include mean field-like theories \cite{approximate, mfprime, mf, mfprimeprimeprime, mfprimeprime, mfprime1,  mf1, cluster, cluster1,Fisher}.

\section{Entanglement Mean Field Theory}

Entanglement mean field theory (EMFT)  was introduced in \cite{emft1} to study the behavior of bipartite entanglement as well as other two-body physical quantities
of many-body systems. Hitherto, 
it has been applied to understand the static properties of such systems \cite{emft1, classicalamader}. Here we apply it to predict the dynamical
properties and in this section, we present the corresponding  consistency equation. 

For definiteness, suppose that the spin model is decribed on a two-dimensional square lattice. 
EMFT begins by assuming that a certain pair of two neighboring spins are ``special''.
Suppose now that the pair \(\{(i,j), (i,j+1)\}\) of the  two-dimensional square lattice is special. 
We will  use the fact that the square of any Pauli matrix is unity. Therefore, the interaction term 
\[
\sigma_x^{i-1, j} \sigma_x^{i,j}
\] 
 in the Hamiltonian in Eq. (\ref{IsingH}) can be written as  
\[\sigma_x^{i-1, j} \sigma_x^{i,j} \sigma_x^{k,l+1} \sigma_x^{k,l+1},
 \]
which is equivalent to
\[
 (\sigma_x^{i-1, j} \sigma_x^{i,j+1}) (\sigma_x^{i,j} \sigma_x^{i,j+1}). 
\]
We now make the entanglement mean field theory approximation by replacing the non-special (operator) pair 
\[ (\sigma_x^{i-1, j} \sigma_x^{i,j+1})
\]
by its mean value
 \( \langle\sigma_x^{i-1, j} \sigma_x^{i,j+1}\rangle /2\). The factor \(1/2\) is due to the fact that only one of two spins in the non-special pair touches the special one. 
Similar approximations can be done for all the other  interaction term in the Hamiltonian. 

The total number of terms in the Hamiltonian that will have a non-trivial
 contribution  to the self-consistency equation, to be written later, is  \(2\nu_E\) where  the ``EMFT coordination number'', \(\nu_E\), is defined  as half
of the number of bonds connecting to the special pair, in the Hamiltonian under consideration. 
As we will see, \(\nu_E\) has a similar significance in EMFT, as the coordination number in MFT.
Therefore, the EMFT-reduced  Hamiltonain in the case of the transverse Ising model in Eq. (\ref{IsingH}) is 
\begin{eqnarray}
H_{TIM}^{EMFT} = -J \nu_E C \sigma_x^{\vec{i}} \sigma_x^{\vec{j}} - \frac{h(t)}{2} (\sigma_z^{\vec{i}} + \sigma_z^{\vec{j}}), 
\label{IsingHemft}
\end{eqnarray}
where  we have assumed that \((\vec{i}, \vec{j})\) is the special pair. There are only two field terms that make non-trivial contributions to the self-consistency equation. 
The factor \(1/2\) in the field terms is again due to the fact that each of them only touches one of the special pair. 
The EMFT-reduced Hamiltonian includes the as-yet-undetermined 
average correlation
\[ C = \langle \sigma_x^{\vec{k}}\sigma_x^{\vec{l}} \rangle,
\]
where \(\vec{k}\) and \(\vec{l}\) are nearby lattice sites. Note that we have not as yet identified the quantum state in which the average is performed.

We will use EMFT to study the properties of the time-evolved state. For definiteness, we assume that the dynamics starts from 
the EMFT ground state, \( |\psi(0)\rangle\), of the system at \(t=0\). Note that the 
Hamiltonian is time-dependent, without which there will be no non-trivial evolution if the system starts off from its ground state.
The EMFT time-evolved state at time \(t\)
is given by
\begin{eqnarray}
 |\psi(t)\rangle = \exp(-i H_{TIM}^{EMFT}(t>0) t/\hbar) |\psi(0)\rangle.
\end{eqnarray}
We now assume that the average correlation \(C\) was performed for
the state \(|\psi (t) \rangle\). Therefore, for consistency, we must have
 \begin{eqnarray}
\label{consistency}
 C = \langle \psi(t) | \sigma_x^{\vec{i}} \sigma_x^{\vec{j}}|\psi(t)\rangle,
\end{eqnarray}
and this is the \emph{dynamical} EMFT self-consistency equation.
Note that it is written in terms of a two-body physical quantity, in contrast to the consistency 
equations in any mean-field-like theory. We would like to emphasize here that this consistency equation is one of the main differences of EMF theory 
with the MF-like theories, where the
self-consistency equation is in terms of single-site physical parameters, like magnetization.  

\section{Entanglement Measure: Logarithmic Negativity}
As mentioned above, EMFT allows us to predict two-body physical parameters, and of them, we will be primarily interested
in nearest-neighbor entanglement of many-body systems. 
There are several measures of entanglement that are known, and for definiteness, 
we consider an entanglement measure, called logarithmic negativity \cite{Vidal-Werner}. 
A quantum state, \(\rho_{AB}\), of   two distinguishable systems \(A\) and \(B\)
is entangled \cite{Werner} if it 
cannot be written as 
\[
 \rho_{AB} = \sum_i p_i \rho^i_A \otimes \rho^i_B, 
\]
where \(\rho^i_A\) and \(\rho^i_B\) are quantum states of \(A\) and \(B\) respectively.

\emph{The Peres-Horodecki criterion:} For two spin-\(1/2\) systems, it is known that a quantum state is entangled if and only if 
 there exists a negative eigenvalue of the partially transposed state \(\rho_{AB}^{T_A}\) \cite{Peres, Horodecki}, where the partial transposition is 
defined as follows. 

Let us express \(\rho_{AB}\) as 
\[
\rho_{AB}= \sum_{i,j} \sum_{\mu, \nu} a_{ij}^{\mu \nu}(|i\rangle\langle j|)_A \otimes (|\mu\rangle\langle \nu|)_B,
\]
where \(\{|i\rangle\}\) and \(\{|\mu\rangle\}\) are respectively orthonormal bases of the systems 
at \(A\) and \(B\) respectively.
Then the partial transposition of \(\rho_{AB}\) is defined as 
\[
\rho_{AB}^{T_A} = \sum_{i,j}\sum_{\mu, \nu} a_{ij}^{\mu \nu}(|j\rangle\langle i|)_A \otimes (|\mu\rangle\langle \nu|)_B.
\]

One can now define a measure of entanglement, called logarithmic negativity, for quantum states of two spin-1/2 systems, as
\[
{\cal L}(\rho_{AB}) = \log_2 (2 N(\rho_{AB}) +1),
\]
where \(N(\rho_{AB})\) is the sum of the moduli of the negative eigenvalues of \(\rho_{AB}^{T_A}\).
Note here that 
the Peres-Horodecki criterion is still a necessary and sufficient criterion for a state to be entangled 
in a system composed of one spin-1/2 and one spin-1, although not any longer 
in higher dimensions \cite{HHHHRMP}. 
Irrespective of that, one can still define the logarithmic negativity for all such situations. However, we will 
not have occasion to consider cases other than two spin-1/2 particles.


\section{EMFT prediction for entanglement of time-evolved state}

It is possible to study single-site as well as two-site properties of the many-body system by solving the 
dynamical EMFT self-consistency equation in (\ref{consistency}). One begins by solving Eq. (\ref{consistency}) for \(C\)
for a certain chosen quantum state of the system, and
inserting it into the EMFT-reduced Hamiltonian. Thereafter, the properties of that state can be obtained by using the EMFT-reduced Hamiltonian, which,
after the insertion of the value of \(C\), 
does not contain any undetermined parameter.

In particular, we can study the behavior of entanglement in the dynamical EMFT state. 
For the quantum TIM described on a lattice of arbitrary dimension and geometry, characterized by the EMFT coordination number 
\(\nu_E\), we find the bipartite entanglement of the EMFT time-evolved state for which the initial state is the EMFT ground state. 
This is therefore to be compared with the nearest neighbor entanglement in the time-evolved state for which the initial state is the ground state of the 
parent Hamiltonian.
In Fig. 1, we plot the bipartite entanglement (as quantified by logarithmic negativity) of the EMFT-reduced quantum transverse Ising model
against the scaled time  \(\tau = t\nu_E J/ \hbar \) 
and the scaled initial transverse field
\(\alpha = a/(\nu_E J)\), when the initial state is the ground state of the EMFT-reduced Hamiltonian. 
\begin{figure}[h!]
\label{fig-chhobi-ek}
\begin{center}
\includegraphics[width=7cm, angle=-90] {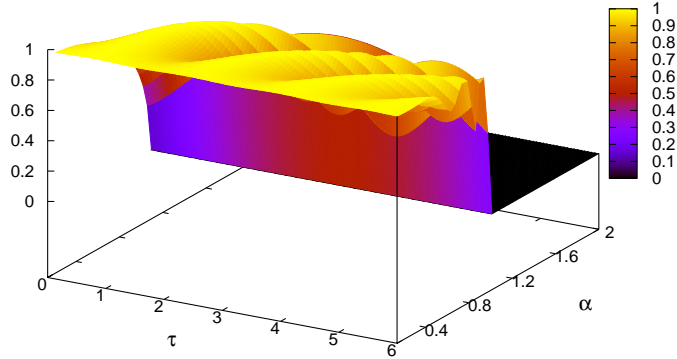}
\caption{(Color online) Behavior of bipartite entanglement of the EMFT time-evolved state, with the initial state of evolution being 
the ground state of the EMFT-reduced quantum transverse Ising model. 
The vertical axis represents entanglement, measured in ebits, while the base axes represent 
the scaled time 
\(\tau = t\nu_E J/\hbar\)  and the scaled initial transverse magnetic field \(\alpha = a/(\nu_E J)\). The parameters of both the base axes are dimensionless.
}
\end{center}
\end{figure}

\section{Predictions for a chain: EMFT versus exact}

The one-dimensional lattice 
%
corresponds to \(\nu_E =1\), and the EMFT prediction for bipartite entanglement in that lattice 
can be read off from the above considerations by substituting \(\nu_E=1\).
As mentioned earlier, this 1D model is exactly solvable by the Jordan-Wigner transformation, 
and it is possible to obtain several single-site as well as two-site physical quantities of the time-dependent Hamiltonian considered here \cite{Barrouch-McCoy}. 
The single-site 
properties include magnetization, while two-site quantities include two-body correlations and entanglement.
The nearest neighbor (bipartite) entanglement of the time-evolved state in this model, as obtained via Jordan-Wigner transformation, and where 
the initial state of the evolution  is the zero-temperature state (of the parent Hamiltonian),
is shown in Fig. 2. Note here that Fig. 2 shows that the bipartite entanglement either vanishes or changes its charateristic (from being convex to concave) at 
\(a/J \approx 1\), irrespective of the time. 
This is expected as we know that this model undergoes a quantum phase transition at zero temperature when the 
(scaled) transverse magnetic field \(a/J\) is unity \cite{Sachdev}.
Compare this with Fig. 1 for \(\nu_E =1\), where we find that the bipartite entanglement as predicted by the EMFT-reduced 
Hamiltonian also  indicates a change of behavior at \(a/J = 1\).
\begin{figure}[h!]
\label{fig-chhobi-dui}
\begin{center}
\includegraphics[width=6cm, angle=-90]{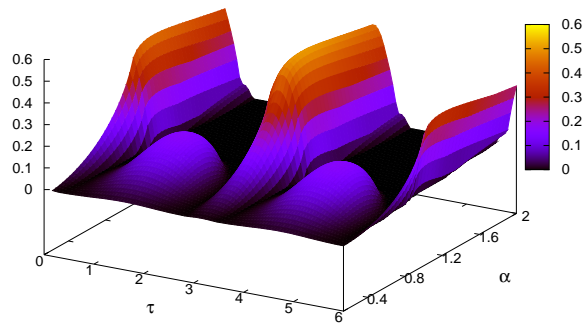}
\caption{(Color online) Nearest-neighbor entanglement of the time-evolved state as obtained in exact calculations (via Jordan-Wigner), with the 
initial state of the evolution being the zero-temperature state of the transverse Ising model.
Entanglement (measured in ebits) is represented along the vertical axis, while the base axes represent the dimensionless variables 
\(\tau=t J/\hbar\) and 
\(\alpha=a/J\). 
Compare 
the entanglement plotted here and that predicted in EMFT in Fig. 1 (for \(\nu_E =1\)).
}
\end{center}
\end{figure}
\vspace{0.5cm}

To make the comparison more objective, 
let us now concentrate on the temporal behavior of entanglement in this model as predicted by EMFT and in the exact case. 
Specifically, we study the logarithmic negativity of the time-evolved state considered above, for a fixed initial disturbance,
with respect to time.  
In Fig. 2, we see that the nearest neighbor entanglement, 
obtained via Jordan-Wigner transformation, shows collapses and revivals with respect to time, if we consider a fixed value of the initial transverse field.
Interestingly, for \(\alpha \leq 1\), the bipartite entanglement of the corresponding EMFT time-evolved state also shows crests and troughs with respect to time
for a fixed initial transverse field.
In Fig. 3, we plot sections of Fig. 1 and Fig. 2 at an exemplary field parameter \(\alpha = 0.594\), and see that the numbers of crests and troughs are equal in the 
two cases.

\vspace{1cm}

\begin{figure}[ht]
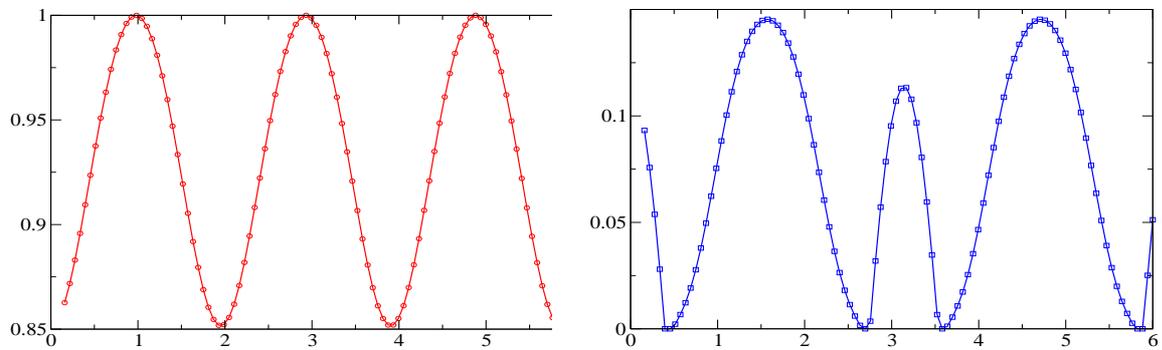

\includegraphics[height=4.5cm, width=7.5cm]{emft_isingdynwitht.eps}
\includegraphics[height=4.5cm, width=7.5cm]{exact_Ising_dynamics.eps}
\caption{\footnotesize One dimensional transverse Ising model: Comparison of temporal behavior of bipartite 
entanglement of the time-evolved state in EMFT with exact results.
The vertical and horizontal axes in both the plots are respectively entanglement (measured in ebits) and 
the scaled time \(\tau=tJ/\hbar\) (dimensionless). 
Both the plots are 
for the fixed initial field 
\(\alpha = a/J = 0.594\). 
The left plot is the EMFT prediction, while the right one is 
the exact result as obtained via Jordan-Wigner transformation.
Both the plots show crests and troughs. 
More quantitatively, the numbers of crests and troughs 
are the same in the two cases. 
}
\label{chobi-tin}
\end{figure}

\section{Conclusion}

Entanglement mean field theory is an approximate method to deal with many-body systems. Its 
%
main difference 
with the 
mean field-like theories is in the self-consistency equations. 
While the self-consistency equations in mean field-like theories deal with single-site parameters 
like
magnetization, those in entanglement mean field theory  deals with single- as well as two-site physical parameters. 
Using the latter, it is possible to
predict two-body correlations as well as entanglement of many-body systems.
While previous studies used entanglement mean field theory to predict properties of static states of many-body systems, 
here we have shown the method to employ it for dynamical states. 
For definiteness, we have considered the paradigmatic  quantum transverse Ising model on a lattice as our many-body system. 
We have shown that the method can 
 predict the behavior of single- as well as two-site properties of the evolved state. In particular, the behavior of entanglement in the dynamics 
could be predicted by using the entanglement mean feld theory.
In the one-dimensional case, we have compared the behavior of two-site entanglement as predicted by EMFT with exact results obtained
via Jordan-Wigner transformation.

\end{document}